# Structural phase transitions and their influence on $Cu^+$ mobility in superionic ferroelastic $Cu_6PS_5I$ single crystals.

A. Gągor[1*], A. Pietraszko[1] M. Drozd[1], M. Połomska[2], D. Kaynts[3]

*Institute of Low Temperatures and Structural Research*
*Polish Academy of Science, Wrocław Poland[1]*
*Institute of Molecular Physic, Polish Academy of Science, Poznań, Poland[2]*
*Uzhhorod State University, Uzhhorod, Ukraine[3]*
Email a.gagor@int.pan.wroc.pl

## Abstract

The structural origin of $Cu^+$ ions conductivity in $Cu_6PS_5I$ single crystals is described in terms of structural phase transitions studied by X-ray diffraction, polarizing microscope and calorimetric measurements. Below the phase transition at $T_c$=(144-169) K $Cu_6PS_5I$ belongs to monoclinic, ferroelastic phase, space group Cc. Above $T_c$ crystal changes the symmetry to cubic superstructure, space group F-43c (a'=19.528); finally at 274K disordering of the $Cu^+$ ions increases the symmetry to F-43m, (a=9.794). The phase transition at 274K coincides well with a strong anomaly in electrical conductivity observed in the Arrhenius plot. Diffusion paths for $Cu^+$ ions are evidenced by means of the atomic displacement factors and split model. Influence of the copper stechiometry on the $T_c$ is also discussed.

**Keywords:** fast-ion conductors, structural phase transitions, X-ray diffraction, argyrodites

[*] Corresponding author.

## Introduction.

The investigations presented in this work are a part of a research concerning studies of ferroelastic, superionic $Cu_6PS_5I_xBr_{1-x}$ crystals which exhibit mixed ionic/electronic conduction. They belong to argyrodite family of compounds with icosahedric structure [1,2]. Among the crystals of this family $Cu_6PS_5Hal$ (Hal= Cl, Br, I) have been recently extensively studied because of possible application in high energy density batteries and sensors likewise the possibility of phase transitions and order-disorder studies [3-8]. Despite the numerous papers related to optical, dielectrical and thermodynamical properties of these compounds the knowledge of the low temperature crystal structures remains very limited. Furthermore, the temperatures of polimorphic phase transition reported by different research groups do not correspond to each other. For example, according to Studenyak et al. [9] at low temperature two phase transitions are observed, a first-order superionic PT $T_s$=(165-175) K and second-order, ferroelastic one at $T_c$=269 K while Girnyk et al. report $T_s$ at 210 K and $T_c$ at 180 K [3]. As far as this crystal is concerned conductivity measurements have been performed only on powder samples. The total conductivity was measured by Khus et al. in the range 293–300 K [1]. It was established that at room temperature conduction is purely ionic ($\sigma_i$~$4x10^{-4}$ at RT), with electronic contribution (hole conductivity $\sigma_h$) less than 1%. At higher temperatures $\sigma_h$ increases and at 423 K becomes equal to $\sigma_i$. The low temperature conduction was measured by Beeken et al. [10]. The strong anomaly with broad and continuous change in activation energy at 270 K was observed on Arrhenius plot and much weaker one around 194 K.

In the present work we analyze the structural changes in $Cu_6PS_5I$ single crystals using X-ray diffraction, domain structure observation and calorimetric investigations. Our main purpose is to correlate the changes in structural disordering of $Cu^+$ mobile ions with the electrical conductivity of this compound.

## Experimental

The $Cu_6PS_5I$ crystals were obtained by the conventional vapor transport method [11] at the Uzhhorod State University. Calorimetric measurements were performed from 130 K to 233 K with Perkin – Elmer DSC-7 instrument with scanning rate 20 K/min. The weight of the sample consisted of small monocrystals obtained under the same growth conditions was 31.447 mg. Ferroelastic domain studies were carried out using a polarizing microscope equipped with Linkam temperature controller, $\Delta T=0.1$ K, for the thin, plate samples oriented in cubic phase perpendicular to <100> direction.

Single-crystal X-ray data collection was performed on Kuma KM-4 CCD diffractometer with CCD camera, ($MoK_\alpha$ radiation, graphite monochromator) equipped with a nitrogen-gas–flow cooling system (Oxford Criosystem Controller), $\Delta T=0.3$ K. To minimize absorption effect a sphere of 0.25 mm diameter was prepared in a compressed air mill and absorption correction for the sphere was applied. The lattice parameters were determined by Bond procedure [12] on a Bond diffractometer ($CuK_\alpha$ radiation, graphite monochromator). Data collection for the sphere sample were performed at 150 K, 235 K, and 293 K. The structure refinement was carried through by direct methods with the SHELXL-97 program package. Positions of $Cu^+$ ions were taken from differential Fourier maps.

## Results and discussion

<u>DSC and domain structure. Temperature and sequence of structural phase transitions.</u>

Differential scanning calorimetry shows that on heating and cooling there are three thermal anomalies connected with the phase transitions with temperatures on cooling run as follows: 274 K, 160 K and 144 K, Fig 1. The large hystheresis in low temperature range and a weaker one in high temperature range indicate first-order phase transitions.

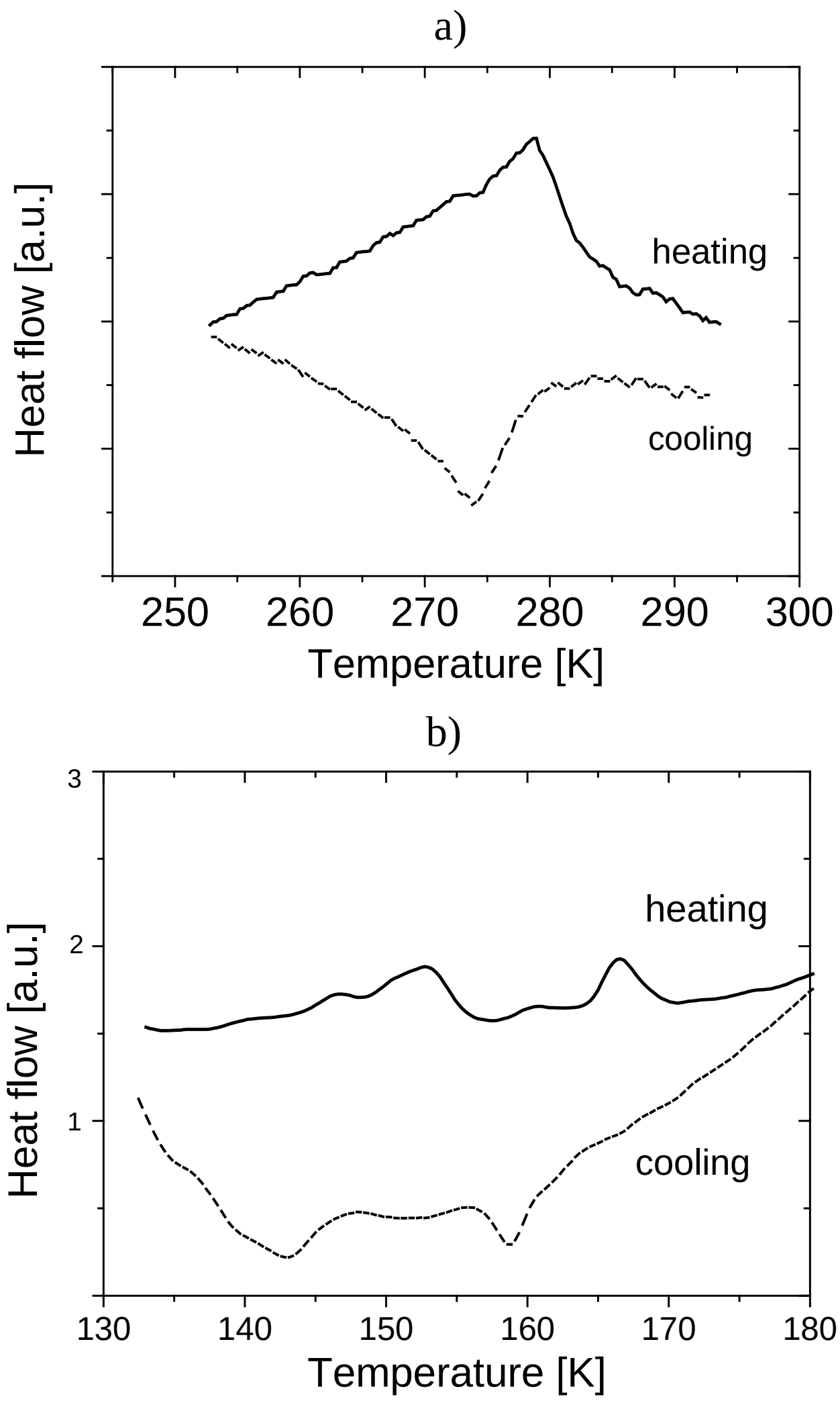

**Fig. 1.** DSC curves of $Cu_6PS_5I$: (a) high, (b) low temperature range.

The investigation of real ferroelastic domain structure and its temperature behavior were performed and compared with DSC measurements. On the basis of this studies the temperature of ferroelastic phase transition was established and the possibility of existence an intermediate phase between high and low temperature phases was checked.

a) b)

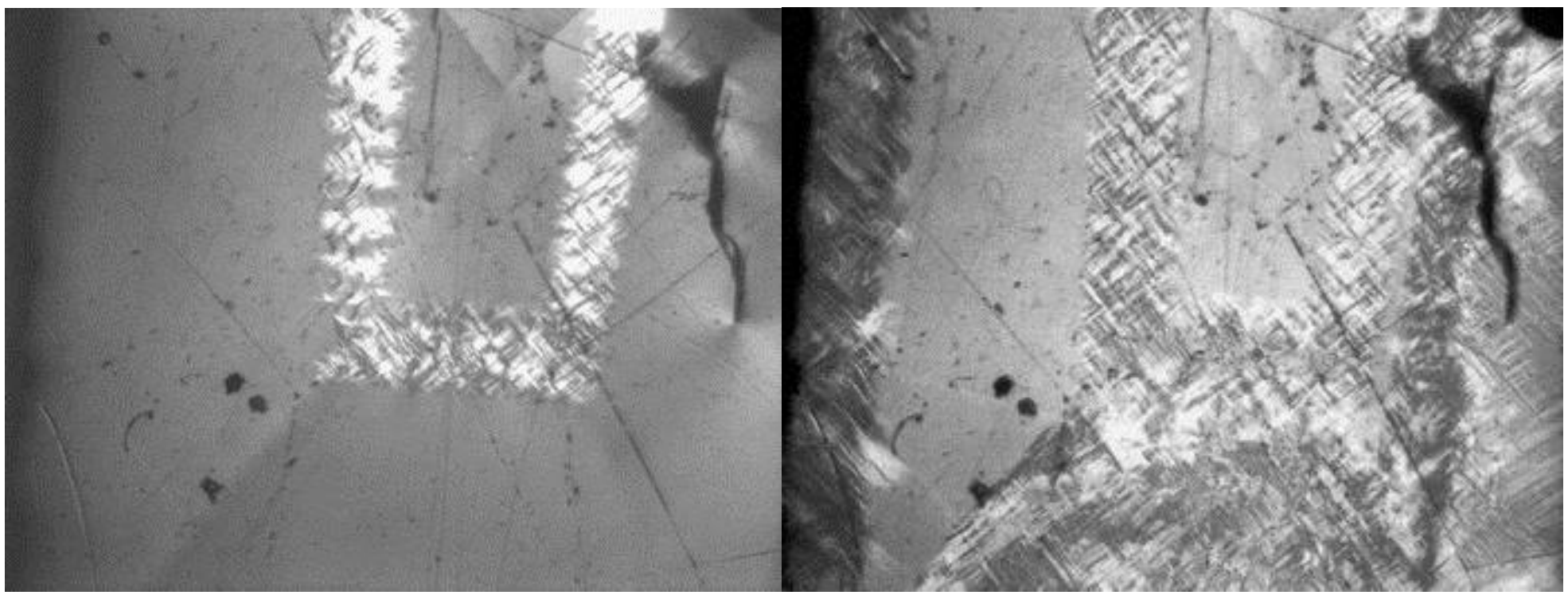

**Fig. 2.** Domain structure in $Cu_6PS_5I$ in <100> direction, (a) 165 K (b) 144 K.

$Cu_6PS_5I$ crystals remain optically isotropic down to 165 K which corresponds to the second peak on DSC diagram. Surprisingly only a part of the sample (rectangularly shaped) undergoes the transition, Fig 2a. The domain structure in the whole of the sample appears below 144 K, Fig. 2b. This transition corresponds to the third peak on the DSC diagram. It was shown by Studenyak [13] that different technological procedures used in a vapor transport method result in synthesis of the crystals with various copper content. This leads to the different temperatures of the superionic phase transition. In our case the samples were prepared under the same growth conditions, therefore it seems that the seeds with different copper content were achieved. The temperature of the ferroelastic phase transition coincides with the temperature $T_s$=169 K of the superionic phase transition based on absorption edge energy reported by Studenyak. It means that in $Cu_6PS_5I$ crystals Tc=Ts, and two phase transitions occur, structural one at 274 K, without changing the crystallographic system and the second, ferroelastic one below 169 K with transition temperature dependent on Cu content in the sample. To confirm achieved results two samples were measured by X-ray diffraction. Precise lattice parameters versus temperature were refined using Bond method for the first sample, splitting of the lattice parameters below 169 K is well pronounced, Fig. 3. In the case the second sample changes in Bragg peak intensities and half widths of the 12 2 2 peak versus temperature were measured, Fig. 4 and 5. Sharp changes in these values occur at 144 K, which supports our interpretation and is in a perfect agreement with results obtained by DSC and domain structure study. The results are summarized on the phase diagram, Fig. 6.

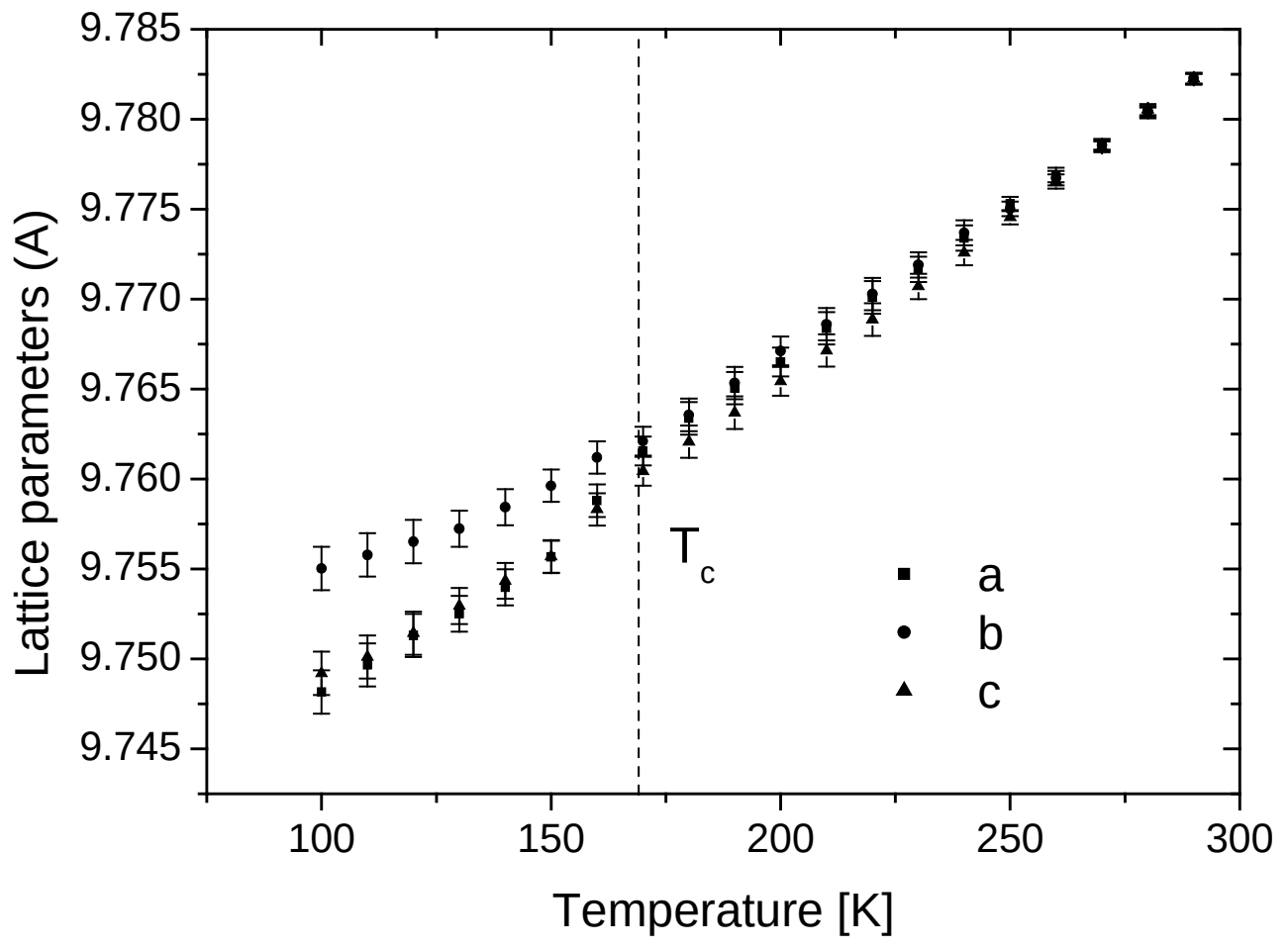

**Fig. 3.** Temperature dependence of lattice parameters *a, b* and *c* [Å].

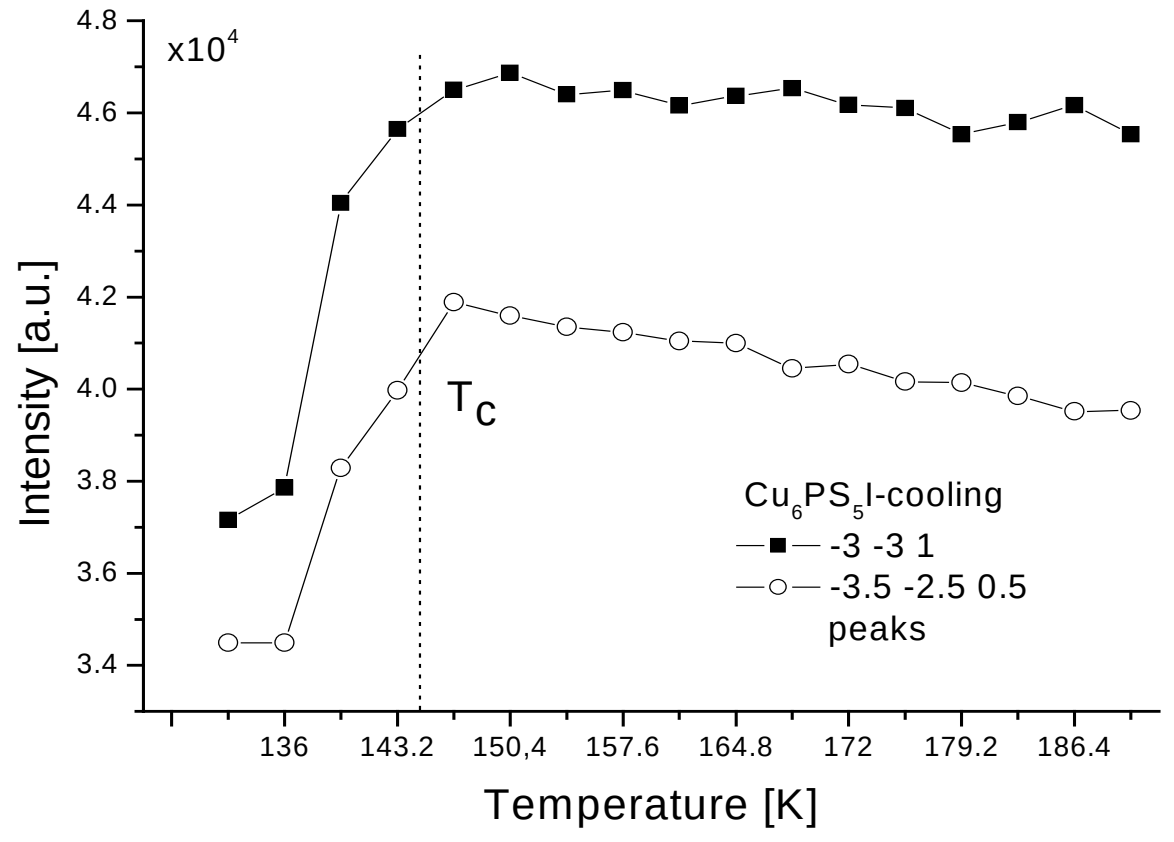

**Fig. 4.** Temperature dependence of intensities of selected reflections indexed in cubic F-43m phase.

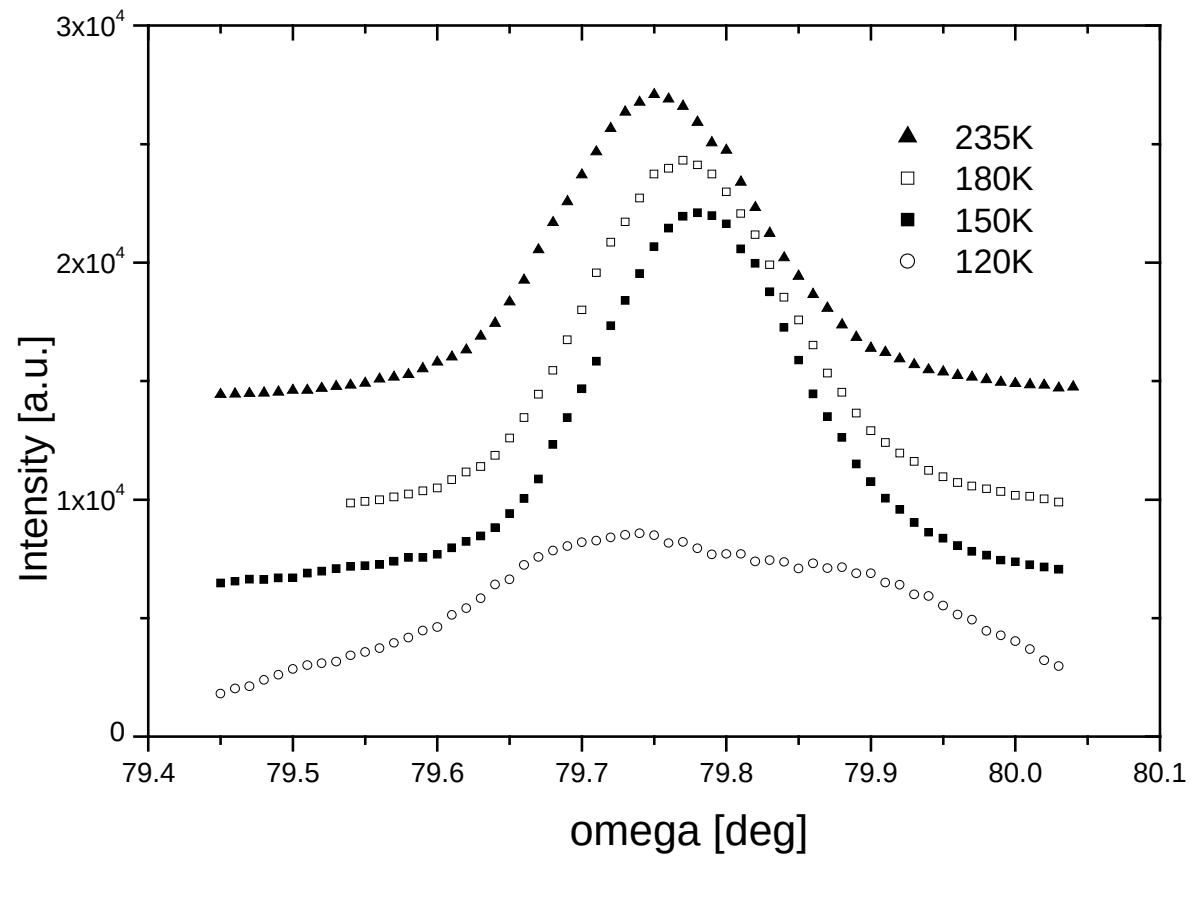

**Fig. 5.** Temperature dependence of the 12 2 2 Bragg peak profile.

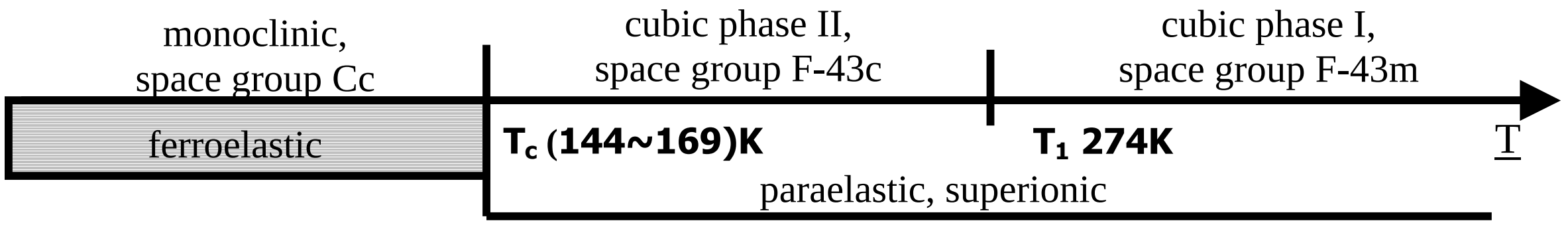

**Fig. 6.** Phase transition diagram of $Cu_6PS_5I$.

Above 274 K $Cu_6PS_5I$ has a cubic structure with F-43m symmetry (cubic phase I) common for all $Cu_6PS_5X$. The extinction conditions reflection data set observed below 274 K is in agreement with F-43c space group with doubled lattice parameters, a=19.528A, (cubic phase II). Below 169 K crystal undergoes ferroelastic and superionic first-order phase transition from cubic to monoclinic system, space group Cc [14]. The temperature of this phase transition depends on the copper content in the sample. Phase transitions in $Cu_6PS_5I$ are connected with ordering of Cu sublattice leading to identical arrangement and full occupation of the Cu sites in all cells in monoclinic phase. The basic icosahedral anion sublattice is preserved after the change of the symmetry.

Because of the low temperature of the ferroelastic phase transition and existence of the cubic phase II between F-43m and Cc symmetry the detailed study of the Cu ordering can be performed in $Cu_6PS_5I$ crystals in the whole range of superionic phase.

$Cu^+$ ionic migration.

In cubic phase I 24 $Cu^+$ ions partly occupy two systems of equivalent tetrahedral sites within a framework of interpenetrating, centered, distorted anion icosahedra. These ions are distributed among 72 positions in unique cell which provides favorable means for ionic migration. $Cu^+$ ions on 48-fold (m) position are coordinated by distorted $S_3I$ tetrahedra and ions which occupy 24-fold (2mm) site are coordinated by planar, triangular $S_3$ configuration, Fig 7a. The change of the symmetry to F-43c at phase II provides six independent 96-fold Cu positions with point symmetry $C_1$ in the rigid skeleton of an anion framework and two different tetrahedral coordinations for $Cu^+$ ions: $I1S_3$ and $I2S_3$.

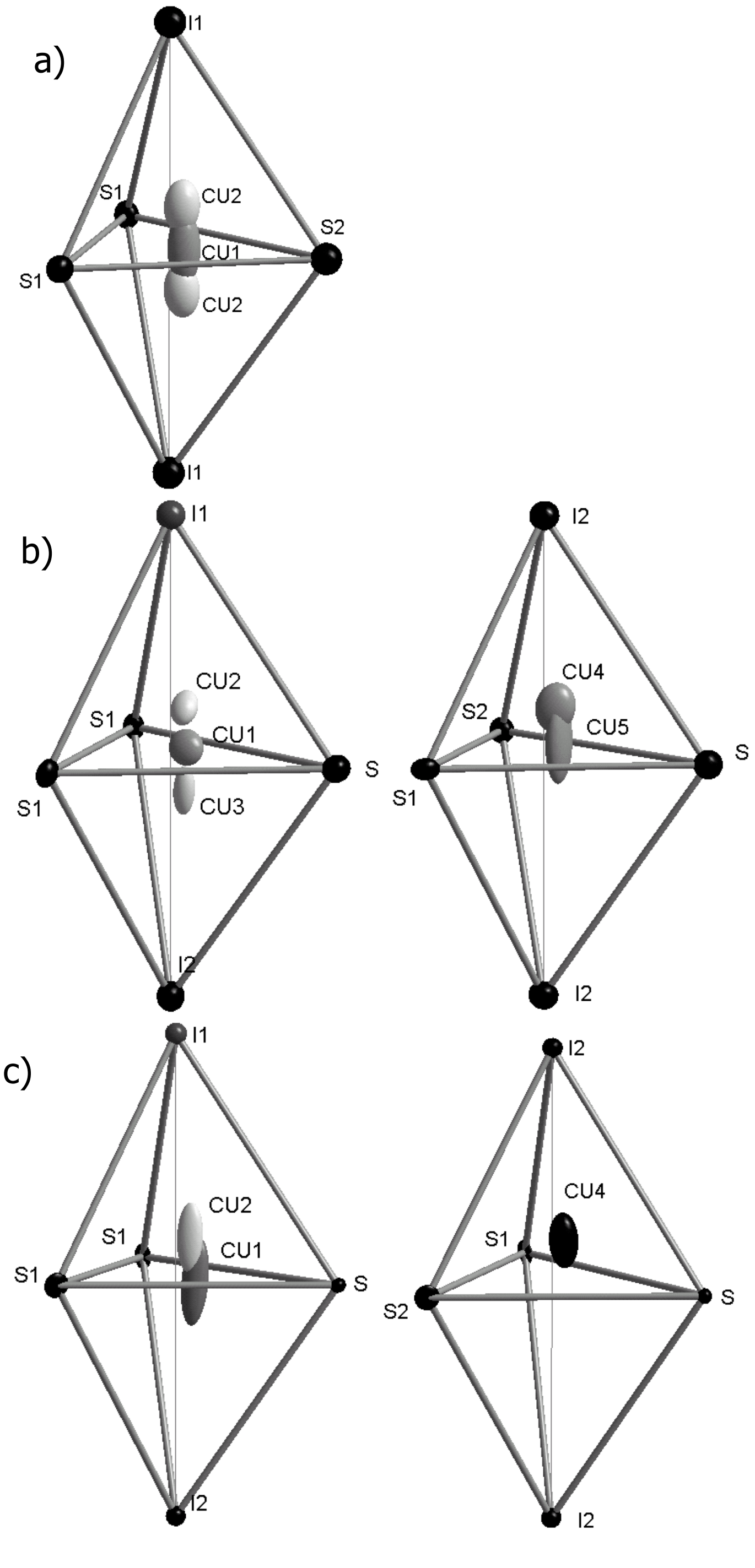

**Fig. 7.** Coordination of $Cu^+$ ions in: a) F-43m phase, T=293 K; b) F-43c phase, T=235 K; c) F-43c phase, T=165 K. The color of ions represents occupancy of the position. Black - full occupancy, dark gray 0.4-0.6, light gray 0.2-0.3.

There are two tendencies in ordering process of the Cu sublattice observed with temperature change. The $I2S_3$ tetrahedral site is strongly preferred by $Cu^+$ ions. During the decrease of the temperature Cu ions shift from planar $S_3$ configuration towards the tetrahedral $I2S_3$ site, Fig. 7b. At 165 K this position is almost fully occupied, Fig. 7c. From the other side ions coordinated by $I1S_3$ tetrahedra prefer the planar triangular $S_3$ configuration. The occupancy of this site increases with temperature lowering but even at 165 K the $Cu^+$ ions are not well localized. Temperature displacement factors are elongated along the axis of the tetrahedra. This strong tendency in occupying triangular planar configuration corresponds to a model of Cu arrangement in monoclinic Cc phase proposed by Haznar et al. in $Cu_6PS_5Br$ compounds [12]. All these structural changes have a great influence on ionic migration. The energy barrier associated with the motion along mobile pathways from the minima at tetrahedral sites to barrier triangular sites increases during ordering of $Cu^+$ ions in tetrahedral sites and also the distances between the Cu atoms elongate thus undoubtedly suppress the mobility of $Cu^+$ ions.

## Conclusions

Structural changes resulted from polymorphic phase transitions have been studied in $Cu_6PS_5I$ single crystals by means of X-ray diffraction, DSC and optical measurements. Below phase transition at $T_c$=(144-169) K crystal belongs to ferroelastic phase with space group Cc. Temperature of $T_c$ depends on copper content in the sample. Above $T_c$ crystal changes the symmetry to cubic superstructure F-43c and at 274 K to F-43m system in structural phase transition. In the F-43c phase $Cu^+$ ions have a tendency to locate in tetragonal and planar triangular sites with lowering of the temperature which suppresses ionic mobility.

## References

[1] W.F. Khus et al. Mat. Res. Bull. 14 (1979) 241-248.

[2] W.F. Khus et al. Mat. Res. Bull 11 (1976) 1115-1124.

[3] I. Girnyk et al. Ukr. J. Phys. Opt. 4, 3 (2003) 147-153.

[4] M. Kranjcec et al. Journal of Physic and Chemistry of Solids 65 (2004) 1015-1020.

[6] I.P. Studenyak et al. Materials Science and Engineering B97 (2003) 34-38.

[7] M. Kranjcec et al. Journal of Physic and Chemistry of Solids 62 (2001) 665- 672.

[8] I.P. Studenyak et al. Mat. Res. Bull. 36 (2001) 123-135.

[9] I.P. Studenyak et al. Akad. Nauk. Ser. Fiz. 56 (1992) 86

[10] R.B. Beeken et al. Journal of Physic and Chemistry of Solids 64 (2003) 1261-1264.

[11] S. Fiechter et al. Journal of Crystal Growth 61 (1983) 275-283.

[12] W.L Bond, Acta Cryst. 13 (1960) 814.

[13] I.P. Studenyak Materials Science and Engineering B52 (1998) 202-207.

[14] A.Haznar, A.Pietraszko, I.P.Studenyak, Solid State Ionics 119 (1999) 31-36